\newcommand{\beqn}{\vspace{-0.25cm}\begin{eqnarray*}}
\newcommand{\eeqn}{\end{eqnarray*}}
\newcommand{\bneqn}{\vspace{-0.25cm}\begin{eqnarray}}
\newcommand{\eneqn}{\end{eqnarray}}
\newcommand{\bracks}[1]{\left[#1\right]}
\newcommand{\parens}[1]{\left(#1\right)}
\newcommand{\expe}[1]{\mathbb{E}\bracks{#1}}

\documentclass[10pt,draft,reqno]{amsart}
     \makeatletter
     \def\section{\@startsection{section}{1}%
     \z@{.7\linespacing\@plus\linespacing}{.5\linespacing}%
     {\bfseries
     \centering
     }}
     \def\@secnumfont{\bfseries}
     \makeatother
\newtheorem{theorem}{Theorem}[section]

\theoremstyle{definition}

\theoremstyle{remark}
\newtheorem{remark}[theorem]{Remark}
\numberwithin{equation}{section}
\begin{document}

\title[Revisiting a Theorem of L.A. Shepp On Optimal Stopping]{Revisiting a Theorem of L.A. Shepp On Optimal Stopping}

\author{Philip A. Ernst}
\address{Philip A. Ernst: Department of Statistics, Rice University, Houston, TX 77005, USA}
\email{philip.ernst@rice.edu}

\author{Larry A. Shepp}
\address{Larry A. Shepp (deceased April 2013); Department of Statistics, The Wharton School of the University of Pennsylvania, Philadelphia, PA 19104, USA}



\begin{abstract}
Using a bondholder who seeks to determine when to sell his bond as our motivating example, we revisit one of Larry Shepp's classical theorems on optimal stopping. We offer a novel proof of Theorem 1 from from \cite{Shepp}.  Our approach is that of guessing the optimal control function and proving its optimality with martingales. Without martingale theory one could hardly prove our guess to be correct. 
\end{abstract}

\maketitle

\section{Introduction}

Consider a
bondholder who must determine when to sell his bond. Each corporate or 
municipal zero coupon bond has a fixed face value at which the bond will be
redeemed at some specific time $b$ in the future and a value at which the
bond is traded and which fluctuates from day to day. Suppose the current price
or value is higher than the face value of the bond. Should the owner of the
bond sell it or hold it for some time, with the hope that the bond will rise to an
even higher value over the face value? 

Brownian motion conditioned to have the current value at time 0 and the face
value at time $b$ is both a simple and mathematically appealing model for the
fluctuations of the price of the bond during $[0,b]$\footnote{For now, we ignore
the fact that this gives an unrealistic model (at least for zero coupon bonds) because
these are observed {\it never} to trade at a price higher than face value.
Most consumers would never buy a zero coupon bond if the price is
higher than the face value. But, if they did, then these models would be useful in setting the optimal selling
price.}
Since only the difference between the current and final values is important, we
may, without loss of generality, assume that the price at time 0 is 0 and is
$-a$ at the final time $b$. The bond holder would like to know for which values of $-a$ and $b$ should he or she
decide to sell his or her security.

Let $W^*_{a,b}(t)$ denote the ordinary Brownian motion process $W(t)$
{\em conditioned} so that $W(0) = 0$ and $W(b) = -a$. We want to choose a
selling time, or optimal stopping time, $\tau$ to maximize
$V^*(a,b) = \expe{W^*(\tau)}$. Suppose we know $V$ and believe in the model, and
that $V(a_0(b),b) = 0$. The result is that if a bond is trading at a price more than $a_0(b)$ 
above its face value $b$ at a time $b$ before the termination date then we
should immediately sell it. \\
\indent Considering optimal stopping problems under a completely different guise (indeed, bond liquidation was far from the preoccupations of the time) the solution to the maximization problem defined in the previous paragraph was solved in \cite{Shepp}! Specifically, in this paper, it is proved that the value of $\alpha $ in the expression $a_0(b) = \alpha \sqrt{b}$, 
is $\alpha = .83992\ldots$. The proof of the latter, according to the authors of \cite{Shepp}, was ``incredibly dense'' because the calculations were made without martingale theory \footnote{Personal communication}. To appreciate how difficult it is to solve the bondholder's problem without using martingale theory, one only need glance at the calculations in \cite{Shepp}. \\
\indent Using the bondholder as our motivating example, we proceed to offer a novel proof of the classical result in \cite{Shepp} by guessing the optimal control function and using martingale theory to prove its optimality. The reader is encouraged to consult \cite{Ernst}, in which the motivation, reasoning, and success of this strategy is well documented.

\section{Guessing the Optimal Control Function}

We first wish to determine for which $a$ and $b$ the inequality $\expe{W^*_{a,b}(\tau)} \le 0$ will hold for all $\tau$. The process
$W^*_{a,b}(t)$ can be written in terms of the ordinary Wiener process
$W(t)$ for $t \ge 0$ by the following simple
formula
\begin{equation}
	W^*_{a,b}(t) = -at/b + (1-t/b) W(t/(1-t/b) ). \label{eq:pinned}
\end{equation}
We justify equation (\ref{eq:pinned}) by first noting that the processes on opposite sides of the equality are the same process. This is because $\sqrt b W(t/\sqrt b)$ is a Wiener process and the process $W^*$ was defined in Section 1 as the Wiener process conditioned so that
$W(1) = 0$ has zero mean and covariance $\text{min}(s,t) - st$. Thus
$W^*$ is the same process as $(1-t)W(t/(1-t) )$.  This allows us to restate the question more simply in terms of $W$ itself.
Indeed, if we make the (monotonic) change of stopping time variable
$\tau = t/(1-t/b)$, then from (\ref{eq:pinned}) we have $t =\tau/(1+\tau/b)$
and
\begin{equation}
	W^*_{a,b}(t) = -a + b \frac{a+W(\tau)}{b+\tau}. \label{eq:dishonest}
\end{equation}
Since $\tau$ runs through all stopping times on $[0,\infty)$ as $t$ runs over
stopping times on $[0,b]$, we see that $a,b$ is a pair satisfying 
$\expe{W^*_{a,b}(t)} \le 0$ if and only if for all stopping times $\tau$ of $W$, we
have

\beqn
V(a,b) := \frac{\expe{a+W(\tau)}}{b+\tau} \le \frac{a}{b}.
\eeqn

We now turn to the problem of finding $V(a,b)$ for all $a,b$. We begin our search
by making a guess, following the same stochastic optimization approach as that of \cite{Ernst}. Intuitively, we should stop at $t = 0$ if $a$ is sufficiently
large compared to $b$, i.e., if $a \ge f(b)$. It is also clear that once $f(b)$
is determined, then the optimal stopping rule will be to stop the first time
$t$ that $a+W(t) \ge f(b+t)$. This is because the problem at time $t$ is the same as
at time 0, with simply a different value of $(a,b)$, namely $(a+W(t),b+t)$.
The unknown function, $a = f(b)$, is the free boundary which we have to find.
It is also clear that if we do quit at $t = 0$, i.e., $a \ge f(b)$, then
$V(a,b) = a/b$. If $a < f(b)$ then we must continue to sample or observe the
ratio, and we must have at some small time $h$ that
\begin{equation}
	V(a,b) \doteq \expe{V(a+W(h),b+h)}.
\end{equation}
We can either stop the process at time 0 or allow the process to run for a 
small time $h$ and reassess. If we do the latter, then, assuming $V(a,b)$ is
twice differentiable in $a$ and once differentiable in $b$, we may expand $V$ in
a Taylor's series, and use It\^o calculus to obtain

\begin{equation}
 V(a,b) = \expe{V(a,b)+V_1(a,b)W(h)+1/2V_{11}(a,b)W^2(h)+V_2(a,b)h+o(h)}.
\end{equation}
\normalsize
Since $\expe{W(h)} = 0$ and $\expe{W^2(h)} = h$, we can subtract $V(a,b)$ and divide by $h$
and let $h \rightarrow 0$. Doing so, we obtain that if $(a,b)$ is a point where we
continue, then we must have the following partial differential equation:
\begin{equation}
	0 = \frac{1}{V_{11}(a,b)} + V_2(a,b). \label{niceone}
\end{equation}
Again, recall that we are still only heuristically just trying to
{\em guess} the right $f$. Since $W$ has the property that the Brownian motion
process scales quadratically in a rescaling of time, i.e.,
$W(t) \sim \sqrt b W'(t/b)$, where $W'$ is another Brownian motion, 
we can write 

\begin{equation}
 V(a,b) = sup_{\tau}\expe{\frac{\frac{a}{b}+\frac{W(\tau)}{b}}{1+\frac{\tau}{b}}} = \frac{1}{\sqrt b} sup_{\tau}\frac{\frac{a}{\sqrt b} + W'(\tau)}{1+\tau} = \frac{1}{\sqrt b} V\parens{\frac{a}{\sqrt b},1},
\end{equation}
as $\tau/b$ runs through all stopping times.
This means that

\beqn
V(a,b) = (1/\sqrt b h)(a/\sqrt b)
\eeqn 
for some $h(x) = V(x,1)$. 
If this is substituted into the partial differential equation above in (\ref{niceone}), we get an ordinary differential equation for the function $h$.
This brings us much closer to our solution. The ordinary differential equation for $h$ is $h^{''}(u)=uh^{'}(u)+h(u)$.
This ordinary differential equation has two linearly independent solutions, $h_1(u) = e^{1/2 u^2}$
and $h_2(u) = \int_0^{\infty} e^{\lambda u - \lambda^2/2} d\lambda$. The latter expression is known as the parabolic cylinder function (the Whittaker function). Every
solution of the ordinary differential equation must be a linear combination of these. The first solution
does not look right for large $u$ since it grows too fast. We discard it
(again, it is our right to do so, as we are only guessing). The other solution is the one we want. The form of $h$
suggests that the free boundary is $f(b) = c \sqrt b$, for some $c$. It remains
to determine the constant multiplier of the second solution to the ordinary differential equation and the
value of $c$. So far, we have
the guess
\begin{equation}
 V(a,b) = B\int_0^{\infty} e^{\lambda a - \lambda^2 b/2} d\lambda, a < c\sqrt b.
\end{equation}
This must fit smoothly to $a/b$ at the boundary $a = c\sqrt b$. This readily
yields two equations (continuity of the zeroth and first derivatives) which
uniquely determine the constants, $B$ and $c$. This gives us the guess, which we
denote $\hat{V}$, as

\bneqn \label{nice}
	\hat{V}(a,b) = (1-\alpha^2)\int_0^{\infty} e^{\lambda a - \lambda^2 b/2} d\lambda, \,\, a < \alpha \sqrt b,
\eneqn
where $\alpha = .83992\ldots$ is the unique root of the transcendental equation
\begin{equation}
 \alpha=(1-\alpha^2)\int_0^{\infty}e^{\lambda \alpha-1/2\lambda^2}d\lambda.
\end{equation}

\section{Proving Our Guess is Correct}
\begin{theorem}
The $\hat{V}$ above in equation (\ref{nice}) gives the correct answer, i.e.,
$V \equiv \hat{V}$. 
\end{theorem}

\begin{proof} 
We use the supermartingale inequality to prove
$V \le \hat{V}$. Define the process
$Y_t = \hat{V}(a+W_t,b+t)$ where $\hat{V}$ is defined as in the above guess.
One can easily check that
$Y$ is expectation-decreasing, $\expe{dY_t} < 0$, if $a > \alpha \sqrt b$ and that
$\expe{dY_t} = 0$ if $a < \alpha \sqrt b$. Thus $Y$ is always
expectation-decreasing, $\expe{dY_t} \le 0$ and so for any stopping time $\tau$ we
have $\expe{Y_{\tau}} \le \expe{Y_0}$. It is also easy to check that $\hat{V}(a,b) \ge a/b$
for all $a$ and $b$, and so we have, for any stopping time $\tau$ that
\begin{equation}
 \expe{\frac{a+W(\tau)}{b+\tau}} \le \expe{\hat{V}(a+W(\tau),b+\tau)} = \expe{Y_{\tau}} \le \expe{Y_0} = \hat{V}(a,b)
\end{equation}
By the definition of $V$ this holds for every $\tau$, and so 
$V(a,b) \le \hat{V}(a,b)$, for all $a$ and $b$. It is easy to see that for the
stopping time $\tau$ defined as the first $t$ for which
$a+W(t) = \alpha \sqrt{b+t}$ that equality holds throughout. Since this is
a legitimate stopping time, we have that $V \equiv \hat{V}$, completing
the proof.
\end{proof}

We thus conclude that the bond should be sold if and only if the current value is at
least $\alpha \sqrt b$ where $b$ is the time until redemption. This is a
reasonable strategy which has been put into practice; see \cite{Boyce}. Further related literature includes \cite{Shiryaev}, \cite{Shepp2}, and \cite{Karatzas}.

\begin{remark}
Another model for bond trading instead of the pinned Brownian
motion model above would be the Black-Scholes model \cite{Black} used analogously 
as before, i.e., pinned exponential Brownian motion. This model 
has the advantage of never taking negative values, which conforms to the reality that bonds do not take on
negative values. However, the (great) disadvantage is that it seems impossible
to obtain the explicit fair value, i.e. to determine when the bond should be
sold. Using (\ref{eq:pinned}), we see
that the problem for pinned exponential Brownian motion amounts to finding

\begin{equation}
	sup_{\tau} \expe{e^{-c \frac{a+W(\tau)}{b+\tau} }}.
\end{equation}
It seems difficult to work with this model analytically, but a numerical solution should valuable. A drawback of this model is that it allows the bond to trade higher than its face value, 
which is almost never observed in reality.
\end{remark}

\section{An Open Question}
We conclude by offering the reader an open question. Consider the process $\hat{W}$ obtained by pinning the
Wiener process to a {\em random} point at time $t = 1$. Condition the
Wiener process by choosing a random variable $X$ and conditioning so that 
$W(1) = X$. Assuming that only the distribution of $X$ is known, but the
sample value is not (although more and more information about $X$ is obtained
by observing more and more of the path of $\hat{W}$), find
$sup_{\tau} \expe{\hat{W}(\tau)}$. This is an unsolved and difficult problem; the
above method of our paper fails and it seems unlikely that an explicit solution can be 
given.

\par\bigskip\noindent
{\bf Acknowledgment.} We thank Professor Frederi Viens for his invaluable suggestions, which greatly helped improve the quality of this manuscript.

\bibliographystyle{amsplain}

\end{document}